\documentclass[12pt,preprint]{aastex}

\newcommand{\rlc}{R_{\rm lc}}

\newcommand\ipl{{\rm IP}_{\rm lag}}

\shorttitle{Reversals of radio emission in B1822-09}
\shortauthors{Dyks, Zhang, Gil}

\begin{document}

\title{Reversals of radio emission direction in PSR B1822-09}

\author{J. Dyks\altaffilmark{1}, Bing Zhang}
\affil{Physics Department, University of Nevada Las Vegas, NV, USA}
\email{jinx@physics.unlv.edu, bzhang@physics.unlv.edu}

\and

\author{J. Gil}
\affil{Physics Department, University of Nevada Las Vegas, NV, USA}
\affil{Institute of Astronomy, University of Zielona G\'{o}ra, Poland}
\email{jag@physics.unlv.edu}

\altaffiltext{1}{On leave from
Nicolaus Copernicus Astronomical Center, Toru{\'n}, Poland}

\begin{abstract}
The pulse profile of pulsar B1822-09 exhibits a very peculiar kind
of mode changing: a ``precursor" appearing just in front of the
Main-Pulse (MP) exhibits periods of nulling, during which an
interpulse (IP) becomes detectable at rotation phase separated by
roughly $180^\circ$ from the precursor.
We propose that this bizarre phenomenon, which requires an
information transfer between the two components, occurs by means
of reversal of a direction of coherent radio emission generated in
the same emission region. 
This interpretation
naturally explains the lack of weak radio emission in the
off-pulse regions, as well as the problem of information transfer
between emission regions associated with the MP precursor and the
IP. The reversals also imply nulling.  
The model has profound physical implications: (i) the
mechanism of coherent radio emission must allow radiation into
two, opposite, intermittently changing directions; 
(ii) the radio waves must be able to propagate through inner regions of the
neutron star magnetosphere with strong magnetic field. Most importantly,
the model implies \emph{inward} radio emission in pulsar magnetosphere.
\end{abstract}

\keywords{pulsars: general --- pulsars: individual (PSR B1822-09)}

\section{Introduction}

The pulse profile of pulsar PSR B1822-09 consists of three
components: a ``precursor" (PR), lagged by only $15^\circ$
by the main pulse (MP), and an interpulse (IP) separated by some
$199^\circ$ from the precursor. In Fig 4b of Gil, Jessner, Kijak, et al.~(1994,
hereafter GJK94)
these components are located at phases $17^\circ$,
$33^\circ$ and $215^\circ$, and are referred to as the leading
component of the MP (called PR in this paper purely for a
convenience of presentation), the trailing component of the MP,
and the IP, respectively. GJK94 revealed a very peculiar and
strong anticorrelation between the intensities of the PR and the
IP: when the precursor enters nulling phase (i.e. becomes
undetectable in single pulses), the interpulse becomes visible in
single pulses and vice-versa. This is dramatically illustrated in
their Fig. 4a, which shows a sequence of sub-integrations
consisting of 10 single pulses (see also Fig.~3 in Fowler \& Wright 1982). 
This behavior is impossible to explain within
existing models. Interpulses separated by nearly 180$^\circ$ from
the main pulses have traditionally been interpreted as the
emission from a magnetic pole on the opposite side of a nearly
orthogonal (NO) rotator, i.e.~a pulsar with the dipole axis
inclined at large angle $\alpha \sim 90^\circ$ with respect to the
rotation axis e.g.~Hankins \& Fowler (1986). 
This idea is contradicted by the apparent
transfer of information between the two poles.
Two other interpretations of
interpulses assume a nearly aligned (NA) geometry, with $\alpha
\sim$ a few degrees. According to one of them, the observer's
line-of-sight cuts through two sides of a narrow hollow beam
(Lyne \& Manchester 1988).
This model seems excluded in B1822-09 because of a
frequency independence of the MP-IP separation
and the lack of extended bridge of emission between MP and IP. In
the second version of NA single pole (SP) model, the line-of-sight
stays between two nested hollow conical beams during most of the
pulse period (Gil 1983; Gil, Kijak \& Seiradakis 1993). 
In this version of SP model the
separation between MP and IP does not vary with frequency, but it
predicts extended bridges of emission at all phases, which are not
observed in B1822-09. 

PSR B1822-09 is therefore a peculiar case which seems to satisfy
properties of nearly orthogonal double pole (NO-DP) model in some
aspects (narrow profile components, no off-pulse emission
components, frequency independence of the MP-IP separation) as
well as properties of nearly aligned single pole (NA-SP) model
(mostly because of the apparent communication between the
precursor of MP and the IP). Thus, (NA-SP) model seems to be
preferred now in the literature (e.g.~Malov, Malofeev, Machabeli, et al.~1997),
despite the apparent problem with lack of the observed bridge of the emission.
However, one should mention that Rankin (1990, 1993) argued in favor
of nearly orthogonal rotator for B1822-09 on independent grounds.

In this paper we propose an interpretation which, however
nonorthodox, solves at the same time both major problems described
above, that is the information transfer and the lack of the
observed off-pulse emission. It also automatically explains why
the separation between the precursor and the interpulse is close,
but not equal to 180$^\circ$, as well as why the phenomenon
\emph{seems} to be so rare and/or unique. 

\section{Interpretation}

Our model assumes that the three components in the pulse profile
of B1822-09 originate from \emph{two} separate regions in the
pulsar magnetosphere, schematically marked as compact sources A
and B in Fig. 1. 
The actual sources may be considerably extended in two directions, 
but nevertheless
they may be perceived as the narrow components in the pulse profile.
Both the precursor and
the interpulse are emitted from region A, whereas the main pulse
(MP) is emitted from region B. The model requires that for this
particular pulsar our line of sight is inclined at the angle
$\zeta$ close to $90^\circ$ with respect to the rotation axis. The
crucial postulate of the model is that the precursor enters the
nulling phase because the direction of radio emission from region A
reverses by 180$^\circ$. Therefore, we can no longer see the
precursor - instead we begin to see the backward emission from
this same region A in a form of the interpulse separated from the
precursor by roughly (but not exactly) half the rotation period
(Fig 1).

Because both PR and IP are emitted from a single emission region
which somehow manages to reverse the radio emission direction,
the information
transfer problem does not arise at all. In principle, the model
can work for both the nearly aligned, and for the nearly
orthogonal rotator. In the NA case, however, the radio emission
region A would have to be located at very large altitudes to be
detectable at $\zeta \sim 90^\circ$ [if one assumes emission
tangent to magnetic field lines and limited to the open field line
region (OFLR), the altitude of region A should be $\la
(2/3)\rlc$]. For lower, conventional emission altitudes from
within the OFLR, the model requires nearly orthogonal geometry
(Fig 1). This is because for small $r$ the impact angle $\beta =
\zeta - \alpha$ is expected to be small, which means that $\alpha
\sim \zeta \approx 90^\circ$. This is in perfect agreement with
the narrowness of the subpulses and with the lack of bridge (or
off-pulse) emission in the profile of B1822-09.

The separation between PR and IP (the latter of which is just a kind of
``backward version" of PR) is \emph{not exactly} equal to 180$^\circ$
because of the effects of
aberration and propagation time delays (hereafter APT effects).
Since the backward emission has a large probability to pass through the
region near the neutron star (NS), the gravitational bending of light 
(GBL, Zhang \& Loeb 2004)
and/or the refraction (RFR, Lyubarsky \& Petrova 1998) 
on the central region of dense plasma surrounding
the neutron star (Michel 1991; Melatos 1997)
may also
affect the phase at which we detect the precursor.
The magnitude of these effects
will depend on the relative arrangement of the observer, the source,
and of the NS.
In the limit of negligible GBL/RFR it can be easily shown that
if the emission region A is located at distance
$\Delta$ from rotation axis, the interpulse (or any
backward component in general)
lags
the precursor (or the corresponding ``directly viewed component")
by $\pi + \ipl$ radians, where
$\ipl=4\Delta/\rlc\ {\rm rad} = 4.8\cdot{10^{-2}}\ {\rm deg}
\ (\Delta/10^6\ {\rm cm})/(P/{\rm s})$, $\rlc = cP/(2\pi)$ is the light
cylinder radius, and $P=0.77$ s is the rotation period of the pulsar.
In the case of negligible GBL/RFR, it holds that $\ipl > 0$, ie the directly
viewed component
is
lagged by the backward component by \emph{more} than half the rotation period.
It is reasonable to assume that in the case of B1822-09 the precursor is the
direct component because it is more pronounced than the IP. As we can see
in Fig 4 of GJK94, the IP indeed lags PR by more than $180^\circ$.

For B1822-09 $\ipl \simeq 19^\circ$ which implies that the emitting region A
is at the distance $\Delta \approx 0.083\rlc \approx 304\cdot10^6$ cm
from the rotation axis. Since we know that the dipole inclination $\alpha$
is large, we get an estimate of the radial distance $r_A$ for
the emission region A: $r_A \simeq \Delta$.
More exactly, $r_A \le \Delta$, which means that $\beta \le 25.3^\circ$
(assuming that the radio emission from A does not extend far beyond the OFLR;
for larger $\beta$ our line of sight would miss the beam).
We can detect both the direct and the backward
component from the same emission region only when $\zeta = \alpha + \beta
\approx 90^\circ$ which implies that $\alpha \approx 90^\circ - \beta \ge
64.7^\circ$. Thus the model implies a nearly orthogonal (or at least ``highly
inclined") rotator with
$\alpha$ in the range between $64.7^\circ$ and $115.3^\circ$.

The question arises what makes this object so special to exhibit
this strange radio emission reversals.
We argue that B1822-09 is not special at all,
and that the phenomenon
most probably occurs in other objects too.
Only in the ``orthogonal viewing" case, ($\zeta \simeq
90^\circ$) do we have the chance to see both the direct and the backward
component \emph{corresponding to the same emission region}.
For $\zeta \ne 90^\circ$ we can only see one of them -- either
the direct one, or the backward one.
So if the phenomenon is universal (or at least not unique for B1822-09)
we should observe spurious components in pulse profiles,
which may be backward components emitted by regions
which we do not observe directly. 

Because the phenomenon leads to appearance/disappearance of
components in pulse profiles, it may be responsible not only for
the mode changing but also for the nulling: the reversal of
emission would be perceived as disappearance of emission for most
viewing angles $\zeta \ne 90^\circ$. 

It is important to note that if the magnetosphere
of B1822$-$09 is symmetrical with respect to the magnetic equator,
then the viewing angle $\zeta$ cannot be exactly 
equal to $90^\circ$. Otherwise we would have observed exactly 
the same emission pattern two times per one rotation period, which is not 
the case for B1822$-$09.
This problem can be remedied in at least two ways:
1) The emission from region A may have the shape of a fan extended
in the $\zeta$ direction. 2) Because of the GBL/RFR, 
we may be located at $\zeta$ slightly different from $90^\circ$, 
and still see the emission after a reversal (see Fig.~2). 
Because of the GBL/RFR, the observer located at $\zeta$ close to,
but different than $90^\circ$ can miss emission from the other
pole. 

\section{Discussion}

The main issue which remains to be resolved is whether the apparent
reversal is caused by the change of the \emph{emission} direction, or just
by the change of the \emph{propagation} direction of radio waves.  
Effects like refraction and reflection of radio waves are widely
considered as a natural phenomenon in pulsar magnetosphere
(eg.~Lyubarsky \& Petrova 1998; Petrova 2000; Fussell \& Luo 2004).
Reflections of the radio waves by an intermittent plasma flow could possibly
explain the peculiar mode changes of B1822$-$09.
However, a reflection by $180^\circ$ is a special phenomenon, which 
requires fine tuning of the propagation direction and the orientation of 
the reflecting surface. It is more natural to assume that the $180^\circ$ flip
is an intrinsic feature of the emission mechanism itself.

Another important issue is that our interpretation implicitly
\emph{assumes} that the jump of PR to the location of IP is caused
by a $\sim 180^\circ$ flip of emission directon. This is obviously
not a unique way to produce an interpulse. Any change of emission direction
from $(\phi, \zeta)$ to $(\phi+\pi, \zeta)$ would result in an interpulse. 
However, such scenario requires that not only the observer must be
located at appropriate viewing angle, but also the angle between the new
and old emission direction must be equal to $2\zeta$. 

Therefore, the $180^\circ$-flip is most natural
from a purely geometrical point of view.
The observed reversal is then an intrinsic
property of the emission region, ie.~the bulk of radiation is emitted into
the opposite direction after the reversal.

\section{Conclusions}

The case of B1822-09 strongly suggests that a physical
mechanism of the coherent radio emission operates as a
two-directional source, 
which intermittently
changes the direction of bulk emission. 
As long as only one emission region is considered,
the phenomenon can have two principal observational consequences:
1) For $\zeta$ close to $90^\circ$ it results in ``false interpulsars",
ie.~objects like B1822$-$09, with interpulses being barely 
``backward images" of other pulse components. 
For low emission altitudes this requires that inner parts of the magnetosphere 
are transparent for the radio waves. 
2) In the other cases ($\zeta \ne 90^\circ$), 
the emission reversal manifests itself as a null. 

Most importantly, the reversal interpretation implies 
that radio waves
in pulsar magnetosphere are sent \emph{inward}, towards the neutron star.
This implication has a large number of consequences 
for
physical, morphological and geometrical studies of pulsars, which so far
have assumed the outward emission only. 
The inward emission hypothesis is elaborated and successfully 
confronted with the
observational data in Dyks et al.~(2005).

\acknowledgments

This work was supported by a research grant at UNLV
and NASA NNG04GD51G (JD \& BZ) and by grant 1 P03D
029 26 of the Polish State Committee for Scientific Research (JG).

\begin{figure}
\epsscale{1.0} 
\plotone{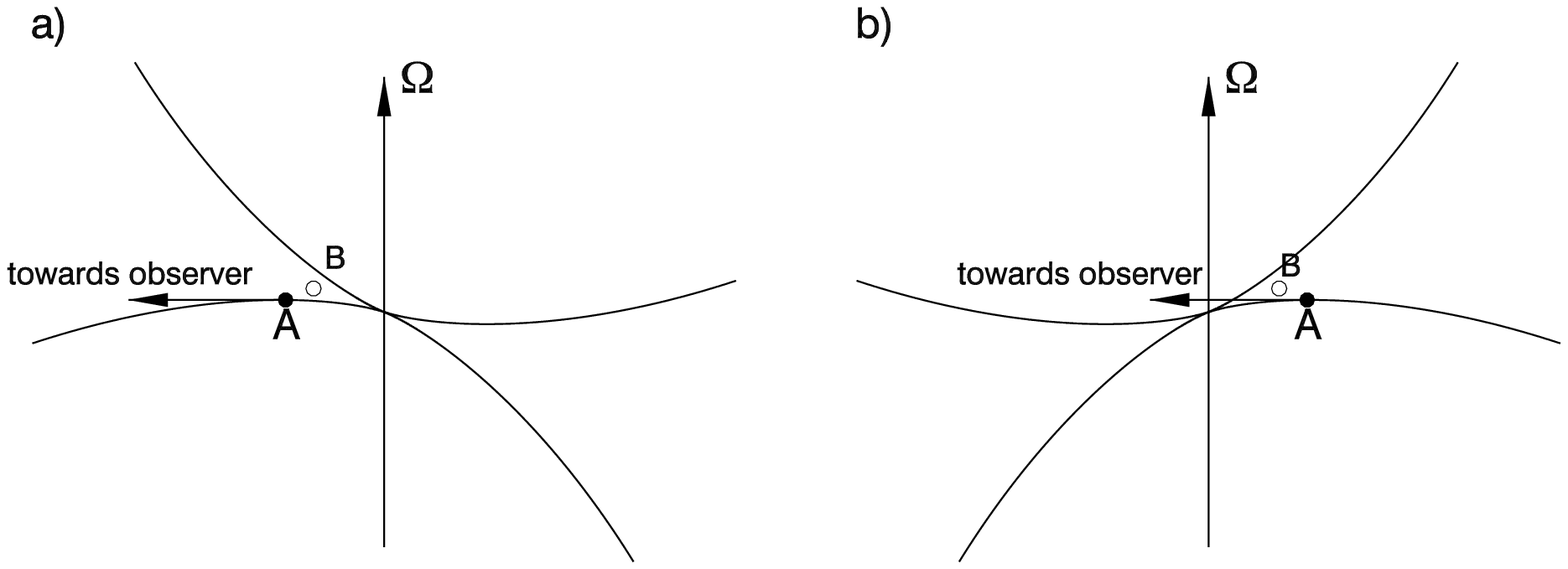} 
\caption{A schematic illustration
of the mechanism of peculiar mode changing observed in PSR
B1822-09. The figures show meridional cut through the inner parts
of the magnetosphere at two instants. The observer is located on
the left side at viewing angle $\zeta \simeq 90^\circ$ (beyond the
figures' area). In (a) radiation from the
source A is emitted outward and is directly received by the
observer at $\zeta \simeq 90^\circ$. As soon as the direction of
emission from A is reversed by $180^\circ$, this source can only
be detected after roughly half the rotation period. The geometry
of emission at that moment is shown in (b).
Radiation from A is now emitted into backward/inward direction,
generally towards the neutron star. The viewing angle does not
have to be exactly equal to $90^\circ$ because the emission beam
from the source A must have some finite angular extent (not
shown).
\label{fig1}}
\end{figure}

\begin{figure}
\epsscale{1.0}
\plotone{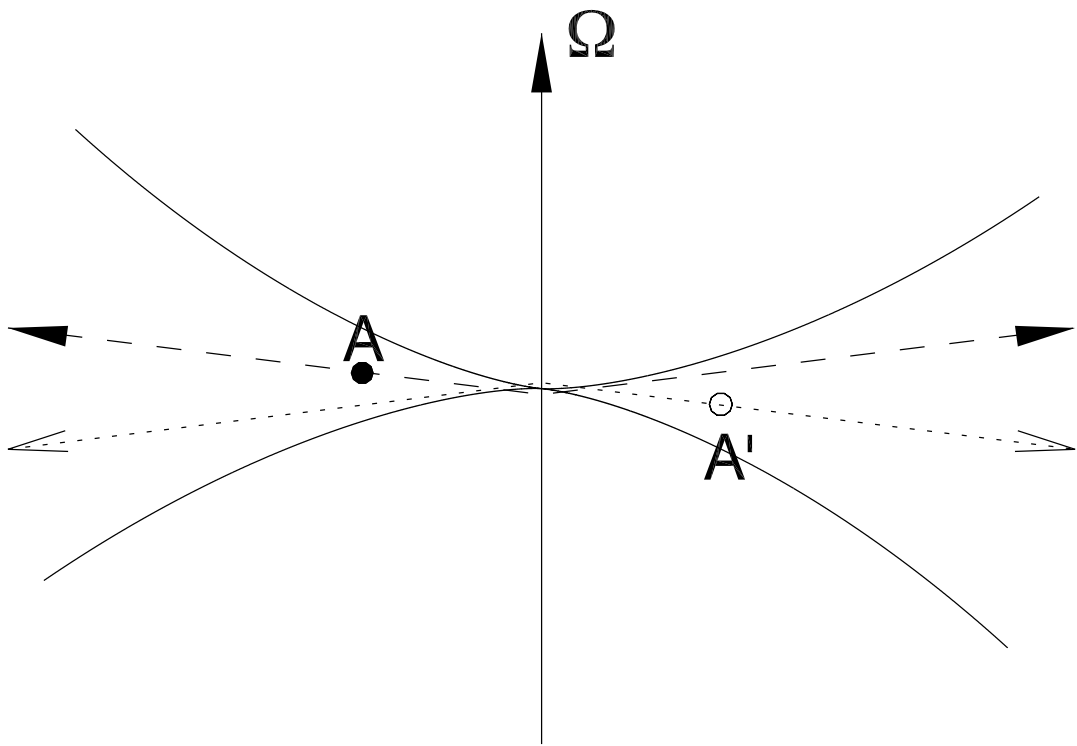}
\caption{Schematic presentation of how the gravitational bending of light
and/or the refraction may affect the reversal model. The purpose of this figure
is to present the possibility, that an observer can view both sides of source A,
but at the same time he can avoid detecting another source A$^\prime$
located
exactly symmetrically on the other side of the neutron star.
The dashed line presents the trajectory
of radiation emitted from A in two opposite directions, tangentially
to the magnetic field lines.
With GBL/RFR included, the observer
can view both the outward, and the inward emission from A
even at $\zeta \ne 90^\circ$ and for an infinitely
narrow beam of emission from A (dashed line).
Radiation from another source A$^\prime$, located symmetrically with respect
to A, may be missed by the observer viewing the source A.
\label{fig2}}
\end{figure}

\end{document}